\RequirePackage{fix-cm}
\documentclass[twocolumn,epjc3]{svjour3}  
\smartqed
\RequirePackage{graphicx}

\RequirePackage{latexsym}
\RequirePackage[numbers,sort&compress]{natbib}
\RequirePackage[colorlinks,citecolor=blue,urlcolor=blue,linkcolor=black]{hyperref}

\usepackage{xspace}
\newcommand{\gerda}{\textsc{Gerda}\xspace}
\newcommand{\majorana}{\textsc{Majorana}\xspace}
\newcommand{\legend}{LEGEND}

\journalname{Eur. Phys. J. C}

\begin{document}

\title{Deep learning based pulse shape discrimination for germanium detectors}


\author{
P.~Holl\thanksref{cr,MPP,INF} \and
L.~Hauertmann\thanksref{MPP} \and
B.~Majorovits\thanksref{MPP} \and
O.~Schulz\thanksref{MPP} \and
M.~Schuster\thanksref{MPP} \and
A.J.~Zsigmond\thanksref{MPP} 
}

\thankstext{cr}{Corresponding author: pholl@mpp.mpg.de}


\institute{Max Planck Institute for Physics, F\"{o}hringer Ring 6, 80805 Munich \label{MPP}
           \and
           \emph{Present Address:} Department of Informatics, Technical University of Munich, Boltzmannstr. 3, 85748 Garching \label{INF}
}

\date{Received: 11 March 2019 / Accepted: 8 April 2019}

\maketitle

\begin{abstract}
Experiments searching for rare processes like neutrinoless double beta decay heavily rely on the identification of background events to reduce their background level and increase their sensitivity.
We present a novel machine learning based method to recognize one of the most abundant classes of background events in these experiments.
By combining a neural network for feature extraction with a smaller classification network, our method can be trained with only a small number of labeled events.
To validate our method, we use signals from a broad-energy germanium detector irradiated with a $^{228}$Th gamma source.
We find that it matches the performance of state-of-the-art algorithms commonly used for this detector type.
However, it requires less tuning and calibration and shows potential to identify certain types of background events missed by other methods.

\end{abstract}

\section{Introduction}
\label{sec:introduction}

Searches for rare processes, like neutrinoless double beta ($0\nu\beta\beta$) decay, critically depend on almost perfect background suppression.
The leading semiconductor based experiments in this field, \gerda~\cite{GerdaPhaseI12} and \majorana~\cite{MajoranaDemonstrator2018}, 
search for $0\nu\beta\beta$ decay of $^{76}$Ge using high-purity germanium (HPGe) detectors.
Extremely low background rates of $10^{-3}$\,cts/(keV$\cdot$kg$\cdot$yr) are required in order to reach their target sensitivity on the $0\nu\beta\beta$ decay half-life.
A significant contribution to their background budget are Compton scattered gamma rays originating from radioactive impurities in the experimental setup~\cite{GERDABackground}.
In many cases, these deposit energy in multiple locations in the germanium detector, producing so-called multi-site (MS) events.
The electrons from $0\nu\beta\beta$ decays, on the other hand, deposit their energy within about 1~mm$^3$ of the detector volume, resulting in single-site (SS) events.
The ability to discriminate between these two event types is therefore crucial.
Experiments planned for the future, such as \legend~\cite{Legend}, will depend even more on the efficiency of background reduction techniques.

Various pulse shape discrimination (PSD) algorithms have been developed in order to identify MS events by analyzing the digitized signal traces of semiconductor detectors in general, and HPGe detectors in particular~\cite{GePSD1993, PSDPhase1, PSDSimulation, Majorana_AoE, Majorana_AoE_2019}.
Due to differences in geometry and electric field configurations, different germanium detector types exhibit different pulse shape characteristics.
Broad-energy germanium (BEGe) detectors are particularly well suited for PSD since over a large fraction of the detector volume, the shape of the current-signal from SS events is almost independent of the location of the energy deposition.
This results in a nearly fixed ratio between peak amplitude, $A$, and integral, $E$, of the current-signal for SS events, but not for MS events.
By using the ratio $A/E$ for PSD, both \gerda~\cite{PSDPhase1} and \majorana~\cite{Majorana_AoE, Majorana_AoE_2019} achieved efficient background rejection for $0\nu\beta\beta$ decay search for BEGe type detectors.
However, the $A/E$ classifier is not able to detect all identifiable MS events since it does not take the full signal shape into account.
It is also not well suited for coaxial HPGe detectors. Therefore, we look for alternative approaches.

Machine learning techniques based on deep neural networks have been used with increasing success for signal processing and analysis in recent years~\cite{EXO-dnn-2018}.
The ability of a neural network to fit a desired function scales primarily with the number of parameters that can be adjusted during training.
However, if the number of trainable parameters is comparable to the amount of training data, a neural network can overfit, i.e. remember the training data without recognizing the underlying patterns.
Large neural networks therefore require large amounts of training data.
In our application, only a small set of labeled training data, i.e. data with SS/MS tagging, is available.
This reflects the typical situation of larger-scale experiments that have to balance calibration and physics data collection time.
It is also difficult to generate labeled data synthetically since the realistic behavior of a full germanium detector system, from charge drift to electronics response, is highly complex and simulations often do not fully describe measured signals.
Therefore, the maximum size of a neural network used for event classification is limited by the amount of labeled data and, consequently, only a small number of input values are possible.

In previous work, this problem was solved by manually extracting a selected number of features from the signals of labeled events~\cite{PSDPhase1,PSDSimulation,BelasThesis}.
By using only these features as input, the artificial neural networks could be built with fewer parameters, at the cost of not being able to make use of all the information contained in the signal.

In this paper, we propose a classification scheme based on two neural networks that are trained independently. While the number of available labeled detector signals is limited, there are ample unlabeled data.
The first network, an autoencoder, is trained on a large part of these unlabeled data in an unsupervised fashion.
It learns to represent the shape of waveforms as low-dimensional feature vectors.
The second network, the classification network, is then trained on labeled data transformed into this feature representation.
This classification network can be built with a small number of trainable parameters due to its low-dimensional input.
Using this two-stage process, all information present in the signal (except for its noise component) can be exploited for event classification, while preventing overfitting.

We first introduce the experimental setup and data taking in section~\ref{sec:setup}. Section~\ref{sec:method} describes the applied PSD technique and section~\ref{sec:verification} its verification. Section~\ref{sec:discussion} provides a comparison with the state-of-the-art $A/E$ PSD method. Conclusions are drawn in section~\ref{sec:conclusions}.

\section{Experimental setup}
\label{sec:setup}

\begin{figure}[ht]
	\includegraphics{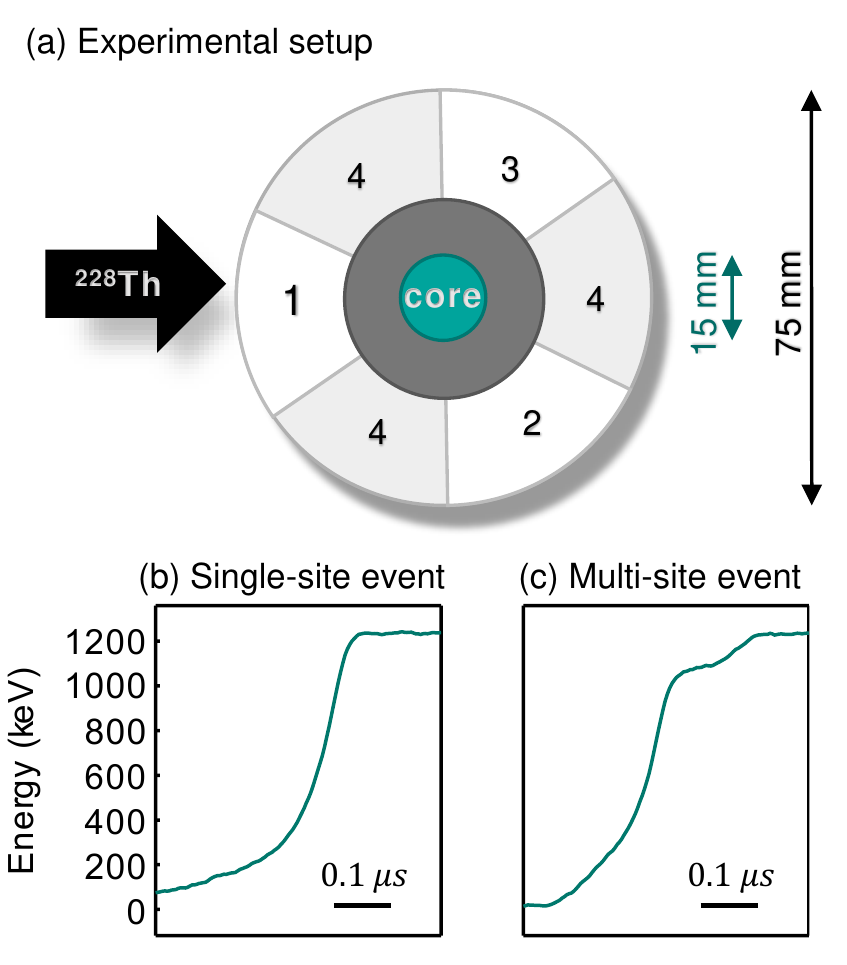}
	\caption{(a) Schematic of the top of the segmented BEGe detector with the n$^+$ electrode (core) and the 4 segments. The ring around the core is passivated. The diameters of the detector and the core contact are 75 mm and 15 mm, respectively. Also shown is the azimuthal position of the $^{228}$Th source. (b,c) The rising part of the charge waveforms recorded by the core electrode from a typical SS and a MS event, both with an energy of 1242~keV.}
	\label{fig:detector}
\end{figure}

We demonstrate our PSD technique using a prototype segmented BEGe detector~\cite{Detector} irradiated with a 1.1~kBq $^{228}$Th source for a period of about 8 hours.
The event count from this exposure is similar to what \gerda and \majorana record in a typical calibration cycle. The radial dimensions of the detector and the position of the source are shown in Fig.~\ref{fig:detector}.
The detector is made from a 40~mm high cylindrical n-type crystal with five readout electrodes: the n$^+$ electrode is called the core and the four p$^+$ surface electrodes the segments.
Segments 1, 2 and 3 cover equally spaced slices of the surface while segment 4 covers the remaining area between them.
A ring around the core contact is passivated.
The detector is enclosed inside a cryostat and the $^{228}$Th source is placed on the cryostat wall at the side of the detector, centered vertically.
The source is centered on segment~1, about 20~mm from the detector surface (see Fig.~\ref{fig:detector}).
A more detailed description of the detector and the setup can be found in~\cite{Detector}.

The signals from the core and segment electrodes are amplified with charge-sensitive amplifiers and digitized with a sampling rate of 250~MHz.
The recorded pulses have a length of $20\mathrm{~\mu s}$ and are centred around the rising edge so that the recorded charge-pulses are long enough for a reliable energy reconstruction.
The deposited energy, $E$, of each event is determined from the total increase in the charge-signal during charge-drift (see Fig.~\ref{fig:detector}b,c). The part of the waveform before the rise is used to determine the baseline level for each event.
To obtain the energy, the baseline level is subtracted from the mean value of the waveform after charge drift, correcting for the decay time of the signal and the cross-talk between segments and core.
The energy is calibrated using the known gamma-line energies of the $^{228}$Th source~\cite{Detector}.
The detector resolution worsens with increasing energy and has a value of around 7 keV and 8 keV (FWHM) for the $^{208}$Tl double-escape peak and $^{212}$Bi full-energy peak, respectively.
The amplitude, $A$, of the current waveform is determined using the procedure described in \cite{PSDPhase1}.

\begin{figure}[t]
	\includegraphics{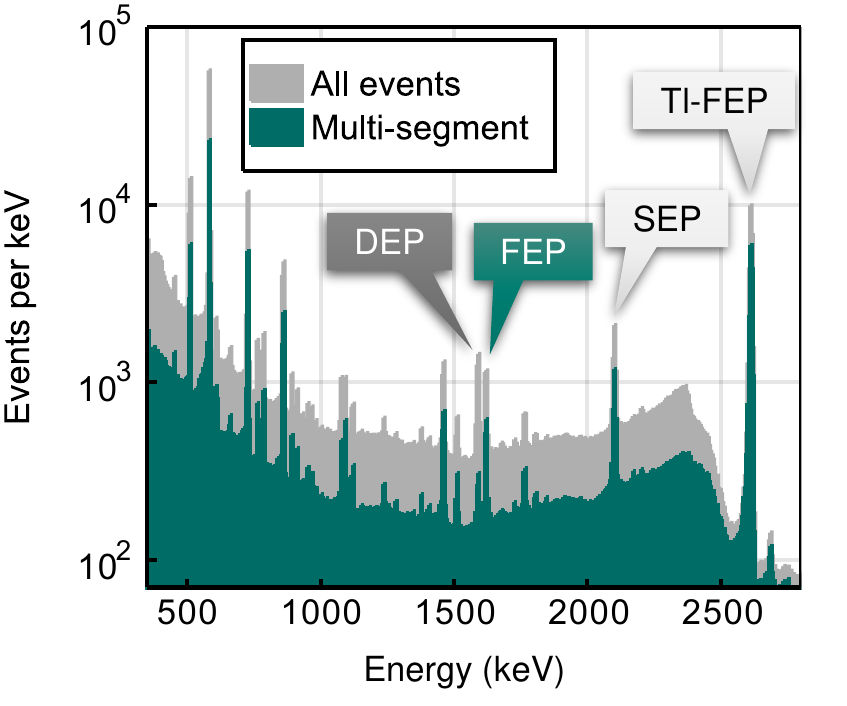}
	\caption{Calibrated energy spectrum from the $^{228}$Th source, reconstructed from the core electrode signals. The gray spectrum shows all events. The green spectrum shows only those events with energy depositions in multiple segments of the detector. The $^{208}$Tl double-escape peak (DEP 1593~keV), the $^{212}$Bi full-energy peak (FEP 1621~keV), the $^{208}$Tl single-escape peak (SEP 2104~keV) and the $^{208}$Tl full-energy peak (Tl-FEP 2615~keV) are marked. Events from the DEP and FEP are used as examples of signal and background events during training of our classifier, the labels are colored accordingly.}
	\label{fig:spectrum}
\end{figure}

Our pulse shape analysis only takes the signals from the core electrode as input. This way, our method is compatible with the hardware deployed in existing $0\nu\beta\beta$ decay experiments, where typically only the core electrode of the germanium detectors is instrumented.
We use the signals recorded from the segment electrodes only to validate the output of our PSD method.

The observed core energy spectrum (Fig.~\ref{fig:spectrum}) shows a Compton continuum, background lines and multiple gamma lines from the $^{228}$Th decay chain, the most prominent of which is the $^{208}$Tl line at 2615~keV.
In the following, we analyze one million events with an energy higher than 1000~keV.

We use the following gamma lines to label events:
\begin{itemize}
    \item the $^{212}$Bi full-energy peak (FEP) at 1621~keV, mostly MS
    \item the $^{208}$Tl single-escape peak (SEP) at 2104~keV, mostly MS
    \item the $^{208}$Tl full-energy peak (Tl-FEP) at 2615~keV, mostly MS
    \item the $^{208}$Tl double-escape peak (DEP) at 1593~keV, predominantly single-site (SS).
\end{itemize}
Events within $\pm4$~keV of one of these peaks are assigned the respective SS / MS label. 
Due to the intrinsic SS / MS mixture of the gamma lines, not all of the assigned labels are correct.
Compton events cause an additional impurity in the labelling because they occur everywhere in the spectrum.
The window size of 8~keV is chosen as a compromise between purity and size of the resulting labeled datasets.
It is comparable to the FWHM of the DEP and FEP while resulting in datasets of similar size.
Table~\ref{tab:datasets} shows the event counts of the labeled datasets as well as the estimated SS fraction in each dataset. The estimated SS fractions are based on Monte Carlo (MC) simulations performed with \textsc{Geant4}.
In the simulations, SS events are defined by $R_{90} < 1$~mm, as in~\cite{Abt:2007iy} and detector resolution as well as Compton events are taken into account.

\begin{table}
\centering
\caption{Overview of the energy-based datasets and associated SS and MS classification including event count and SS fraction with statistical uncertainty obtained from MC simulation.}
\label{tab:datasets}
\begin{tabular}{llll}
\hline\noalign{\smallskip}
Dataset & Event count & SS (\%) & Label  \\
\noalign{\smallskip}\hline\noalign{\smallskip}
DEP $\pm$ 4~keV    & 10.8 k & $85.5 \pm 0.3$ & SS \\
FEP $\pm$ 4~keV    & 10.0 k & $20.1 \pm 0.4$ & MS \\
SEP $\pm$ 4~keV    & 15.0 k & $12.8 \pm 0.2$ & MS \\
Tl-FEP $\pm$ 4~keV & 73.4 k & $6.3 \pm 0.1$ & MS \\
\noalign{\smallskip}\hline
\end{tabular}
\end{table}

In addition to the SS / MS labels, the segmentation of the detector allows us to label events as either single-segment -- events which deposit energy only in segment~1 -- or multi-segment -- those which additionally deposit energy in another segment.
These labels can be used as alternative approximations for SS and MS event labels. They are neither used in network training nor for filtering the training datasets. Instead, they serve to independently verify the classification outputs of our method (see section~\ref{sec:discussion}).

\section{Method}
\label{sec:method}

We perform a number of signal processing steps on the raw waveforms before passing them to the neural networks for further analysis.
The neural network analysis consists of two stages:
In the first stage, the autoencoder extracts features from all preprocessed waveforms of unlabeled events and stores them in a low-dimensional feature vector.
The feature vectors of labeled events are then passed to the classifier network in the second stage.

To train both networks, the total dataset is split into 60\% for training and 40\% for testing. Both the autoencoder and classifier networks are then trained and evaluated on subsets of these two datasets.

\subsection{Preprocessing}
\label{sec:preprocessing}

\begin{figure}
	\includegraphics{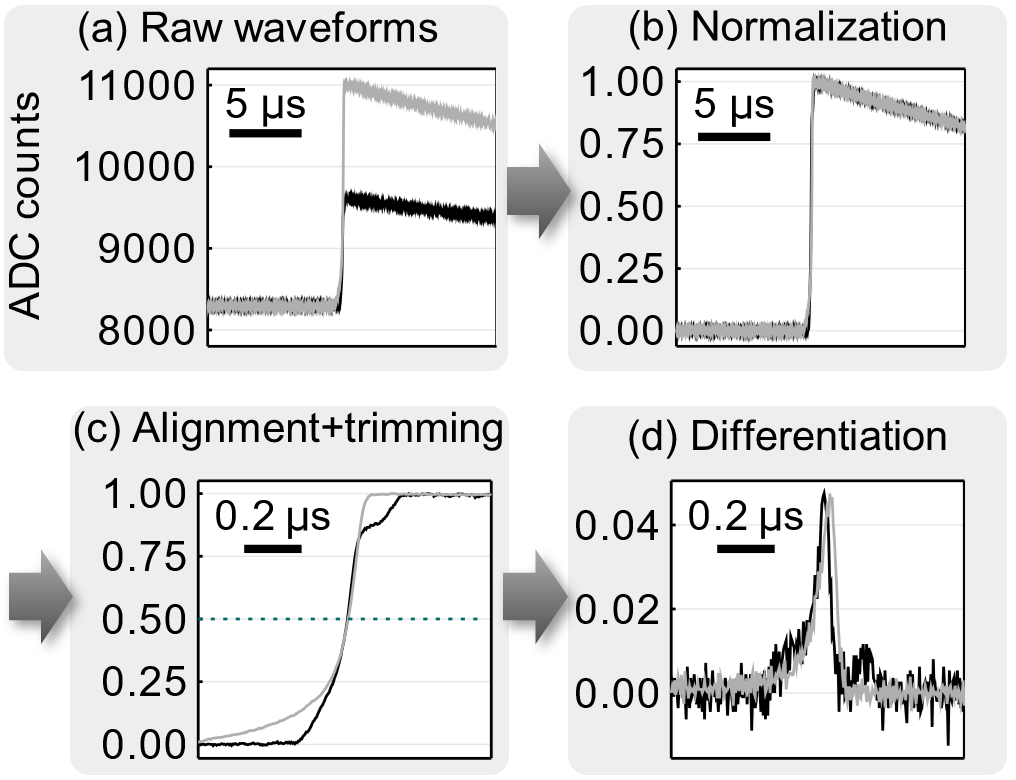}
	\caption{Preprocessing steps shown for the MS event from Fig.~\ref{fig:detector}b (black curve) and a presumed SS event with 2578~keV energy (gray curve). The raw waveforms (a) are baseline-subtracted and normalized to a charge of one (b), aligned at the intersection with a threshold of 0.5, then trimmed to a 1~$\mu$s long window (c), and finally differentiated yielding the current-signal (d).} 
	\label{fig:preprocessing}
\end{figure}

The raw charge-waveforms, digitized from the core electrode, span a time of 20~$\mu$s (Fig.~\ref{fig:preprocessing}a).
In the first preprocessing step, the baseline is subtracted and the waveforms are normalized to their total charge (Fig.~\ref{fig:preprocessing}b).
The normalized waveforms are then aligned in time so that all of them reach a value of 0.5 at the same sample number. We then trim the waveforms to a symmetric 1~$\mu$s window around the alignment point (Fig.~\ref{fig:preprocessing}c).
The aligned and trimmed waveforms consist of 256 samples that cover the entire rise of the signal.
Finally, a differentiation step is performed to obtain the current-waveform from the charge-waveform (Fig~\ref{fig:preprocessing}d) where individual peaks correspond to spatially separated energy depositions in the detector.
Thus, events with one distinguishable peak are assumed to be SS events while events with multiple peaks are MS events.

\subsection{Autoencoder}
\label{sec:autoencoder}

\begin{figure}
	\includegraphics{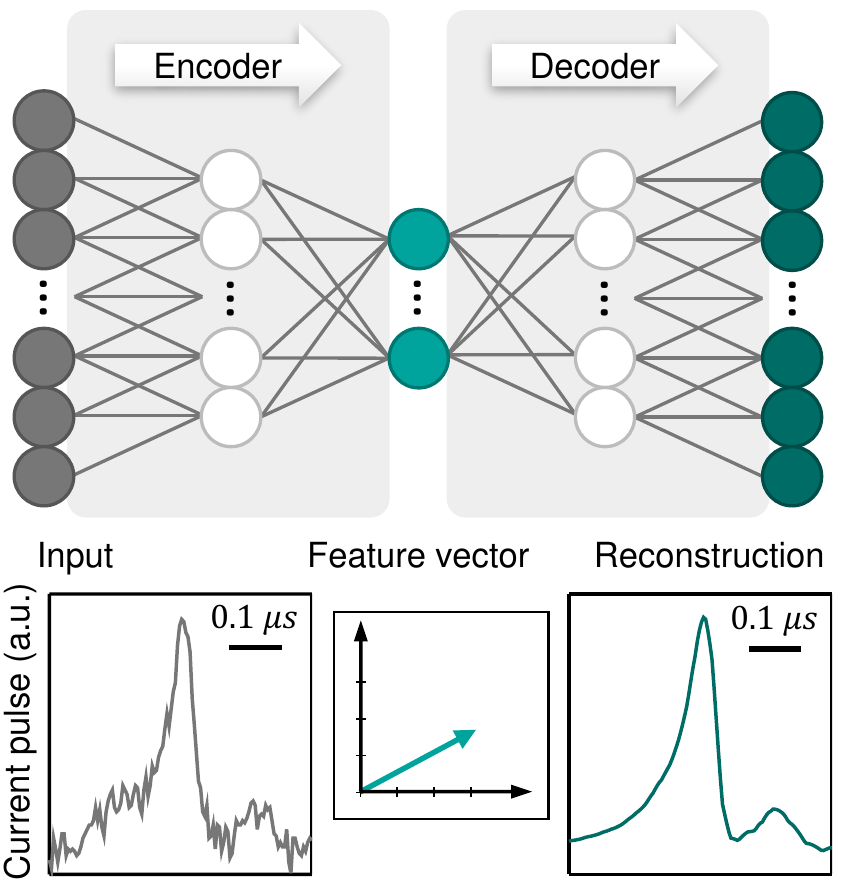}
	\caption{Working principle of the autoencoder network. The symmetric network first encodes an input current-signal (left) into a low-dimensional feature vector (center) before decoding, i.e. reconstructing the input from the feature vector (right).}
	\label{fig:autoencoder}
\end{figure}

After preprocessing, the resulting current-waveforms are used as input to the autoencoder.
The autoencoder is a convolutional neural network.
Its layout consists of two parts, shown in Fig.~\ref{fig:autoencoder}: the encoder extracts the important features from the waveform, storing them in a low-dimensional feature vector, and the decoder attempts to reconstruct the waveform from the feature vector.

In the encoder, a convolutional layer first performs two convolutions with trainable filters on the 256-dimensional input vector, producing two vectors of the same length. Both filters have a length of nine samples plus a constant bias.
The convolutional layer is followed by an activation, applying a rectifying linear unit, $\mathrm{ReLU}(x) \equiv \max(0,x)$, to each value $x$ of its input.
Next, a pooling operation quarters the time-resolution from 256 to 64 by picking the maximum value of each 4 neighbouring samples across the vectors.
A fully connected layer then transforms the reduced vectors into a low-dimensional vector. This operation is implemented as a matrix multiplication where all entries of the $(2 \cdot 64 \times \textrm{feature vector dimension})$ matrix are trainable. Another ReLU is applied to produce the feature vector.
The use of convolution, activation and pooling has become common in computer vision research, where 2D convolutions on images are employed instead of 1D temporal convolutions~\cite{ObjectDetectionReview}.
Since key operations of the encoder depend on trainable parameters, the encoding step is flexible and can map a wide variety of possible functions. All trainable parameters are randomly initialized before training.
It is not possible to predict what information each individual entry of the feature vector will represent after training, and there is no obvious interpretation of the feature representation.

Trials established that seven parameters in the feature vector are sufficient as input to a lightweight classifier network.
Seven parameters are also enough to ensure that all waveforms are reconstructed with sufficient accuracy and that the training converges reliably.
The encoder, with only two hidden layers, proves to be powerful enough to capture the underlying structure of the waveforms, as will be discussed in section~\ref{sec:verification}.

The layout of the decoder mirrors the encoder: it consists of a fully connected layer followed by a ReLU activation, a four times upsampling operation and a deconvolution.
The goal of the decoder during training is to reconstruct the original waveform from the feature vector.
The mean squared error (MSE) is used to measure the accuracy of the reconstruction and as the loss function to be minimized during training:
\begin{equation}
L_\textrm{MSE} = \frac{1}{2NM} \sum_{n \in \mathcal D} \sum_{i \in \mathcal S} \left( x_{n,i} - x_{n,i}^* \right)^2,
\label{eq:MSE}
\end{equation}
where, $\mathcal D$ denotes the training data containing $N$ events, $\mathcal S$ the set of the $M=256$ sample indices of each waveform and $x$ and $x^*$ represent a value of the reconstructed and the original waveform, respectively.
Both encoder and decoder are trained together as a single network that tries to reproduce the input waveform.
This way, the encoder learns to extract the information from the waveform that yields the most faithful reconstruction, focusing on the underlying structure rather than the noise of the signal.

In principle, all recorded events could be used to train the autoencoder since no labels are required.
However, we discard events with energies lower than 1000~keV as they are less relevant for $0\nu\beta\beta$ decay searches and are more affected by noise.

Because of their more complex pulse shapes, MS events require more information to model than SS events.
However, our dataset contains more similar-looking SS events than MS events.
To counteract this, we drop a fraction of SS events to balance the datasets, leaving 725k events.
This filter is based on the $A/E$ value and drops a large fraction of SS events which look almost identical except for noise.
Discarded events are chosen randomly to prevent introduction of a bias.

Examples of current waveforms and their reconstructions are shown in Figs.~\ref{fig:autoencoder} and \ref{fig:disagreement}.
The reconstructions exhibit the same shape as the original waveforms but lack the high-frequency noise due to its high entropy.
A quantitative analysis of the reconstruction quality is presented in section~\ref{sec:verification}.

\subsection{Classifier}
\label{sec:classifier}

The training of the autoencoder is followed by the training of the classifier network.
The DEP and FEP events serve as the SS and MS training datasets, respectively (see section~\ref{sec:setup}).
Their small difference in energy ensures that the noise level is very similar for the two peaks, an additional safeguard that prevents the training to be influenced by varying signal-to-noise ratios.

\begin{figure}
	\includegraphics{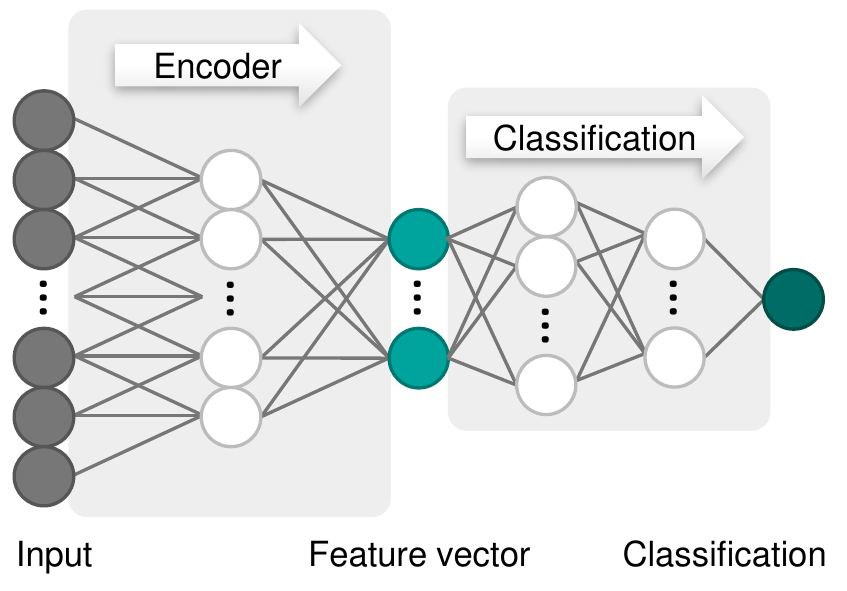}
	\caption{Architecture of the combined network consisting of the encoder to extract feature vectors followed by the smaller classification network to identify events as SS or MS.}
	\label{fig:classifier}
\end{figure}

The classification network takes the low-dimensional feature vector as input.
Two fully connected layers consisting of 10 and 5 neurons with ReLU activation functions process the feature vector and an output layer produces a single value, $c \in (0,1)$, which is correlated to the probability of a given event to be SS (see Fig.~\ref{fig:classifier}).

With the described network architecture, the classifier has a total of 141 trainable parameters.
The training set contains around 30 times as many events per class to ensure that the classifier cannot overfit and remember individual events from the training dataset.
This lightweight network architecture is only possible because the underlying structure of the raw waveforms has already been extracted by the autoencoder.
Again, we use MSE loss (see equation~\ref{eq:MSE}) for training but adjust only the parameters of the classifier network, leaving the previously trained autoencoder unchanged.

\begin{figure}
	\includegraphics{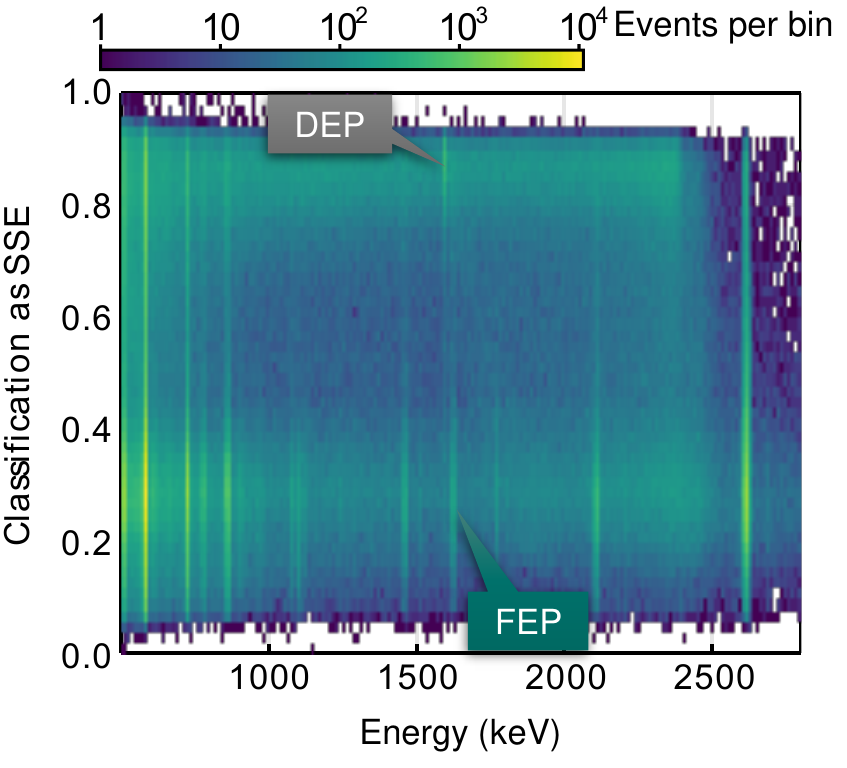}
    \caption{Distribution of the output, $c$, of the combined encoder+classifier network for the test dataset (bin size X:~10~keV, Y:~0.02). The autoencoder is trained on events above 1000 keV and the classifier on events from the DEP (mostly SS) and FEP (mostly MS). The DEP events cluster around 0.9 while the peaks containing mostly MS events are centered at about 0.3.}
	\label{fig:classification-distribution}
\end{figure}

The output values of the classifier are shown in Fig.~\ref{fig:classification-distribution} for the complete test dataset. They demonstrate that the peaks are classified as expected: DEP events are clustered around 0.9, while events from MS peaks cluster below 0.5.

\section{Verification}
\label{sec:verification}

We verify our method using the test dataset, which has 405k events above 1000~keV.
Out of these, 232k events deposited energy in segment 1 and are therefore labeled as single-segment or multi-segment (see section~\ref{sec:setup}).
First, the reconstruction accuracy of the autoencoder is examined on the whole test dataset before the discrimination performance of the combination of encoder and classifier (E+C) is evaluated on events with single-segment/multi-segment labeling.

The autoencoder is trained to keep as much relevant information of the waveform as possible when encoding.
To assess its performance, we define the reconstruction error, $\varepsilon_n$, of an event $n$ as the normalized RMS difference between the original and reconstructed waveform
\begin{equation}
\varepsilon_n \equiv \frac 1 {\sigma_n} \sqrt{\sum_{i\in \mathcal S} (x_{n,i}-x_{n,i}^*)^2}.
\label{eq:error}
\end{equation}
Here, $\sigma_n$ denotes the noise level of the normalized waveform from event $n$, so $\varepsilon$ is constructed to be independent of this noise.
A small $\varepsilon$ thus indicates that much of the information from the original pulse is contained in the reconstruction.
For $\varepsilon = 1$, the reconstruction error is equivalent to the deviation expected from noise only.

\begin{figure}
	\includegraphics{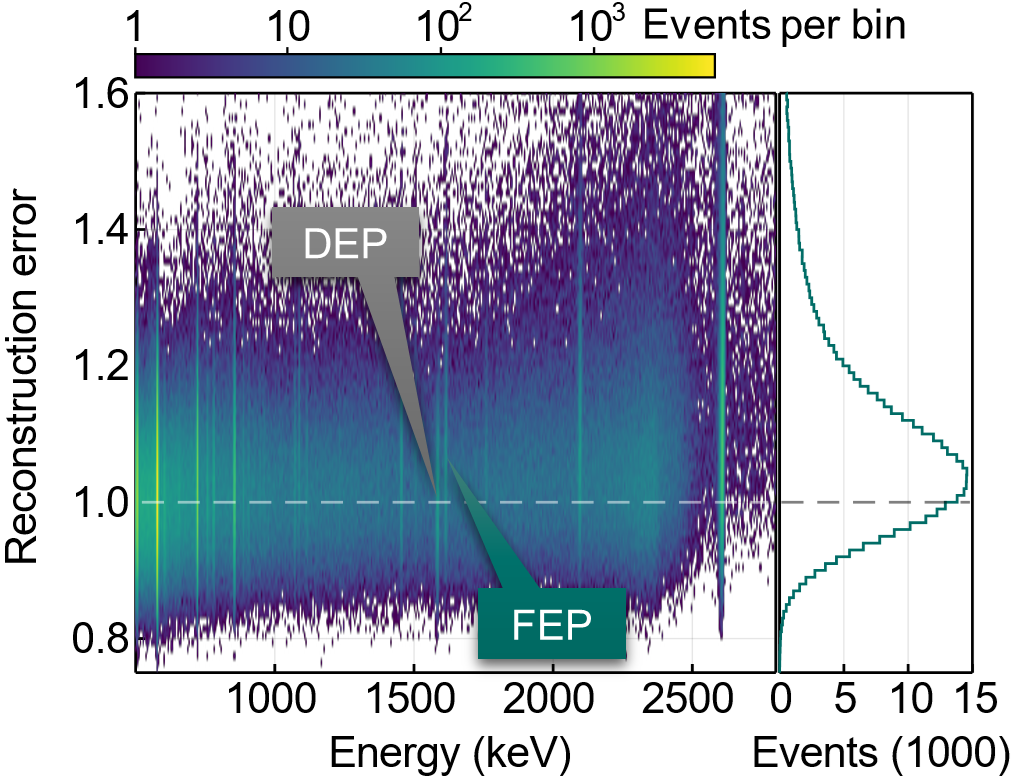}
	\caption{Distribution of the normalized reconstruction error, $\varepsilon$, of the autoencoder as a function of energy (left) and its marginalized distribution for events with an energy between 1000 and 3000 keV (right).}
	\label{fig:reconstruction-accuracy}
\end{figure}

Figure~\ref{fig:reconstruction-accuracy} shows the distribution of $\varepsilon$ as a function of energy.
Single-site events are reconstructed with high accuracy since their consistent shape makes them easy to learn.
As a result, SS events from the DEP form a Gaussian-like distribution around $\varepsilon \approx 1$.
Multi-site events, on the other hand, are more difficult to reconstruct since they comprise events with multiple peaks in the current waveform.
Therefore, the gamma lines dominated by MS events, like the FEP, contain more events with less accurate reconstructions.

Generally though, $\varepsilon$ is small and even complex signal structures of MS or rare outlier events are usually reconstructed well, indicating that the feature vector encodes all important waveform characteristics.
In addition, studies of similar network architectures have shown that the information loss of autoencoder reconstructions can primarily be attributed to the decoder part of the network \cite{ImagesAutoencoderReconstruction}, which is not used for our classification.
For these reasons, the extracted feature vectors constitute an ideal basis for the second classification stage and classification performance is not affected by the information loss.

To evaluate the performance of the classifier, we introduce a variable discrimination threshold.
Events above the threshold are classified as SS events and accepted, while the ones below are classified as MS events (background) and rejected.
Varying this threshold results in different survival fractions for each gamma peak.
The SS fractions of the test datasets resulting from a specific threshold are detailed in table~\ref{tab:peakclasses}.

\begin{table}
\centering
\caption{Classification of events in the test datasets. The class assignment is based on the output, $c$, of the combined encoder+classifier network. Events are counted as SS if $c>0.6$ (maximum class separation threshold). The given uncertainties are statistical only.}
\label{tab:peakclasses}
\begin{tabular}{lll}
\hline\noalign{\smallskip}
Dataset & Event count & Classified SS (\%)  \\
\noalign{\smallskip}\hline\noalign{\smallskip}
DEP $\pm$ 4~keV & 4.3 k  & $82.8 \pm 1.4$ \\
FEP $\pm$ 4~keV & 4.1 k  & $28.4 \pm 0.8$ \\
SEP $\pm$ 4~keV & 6.0 k  & $21.7 \pm 0.6$ \\
Tl-FEP $\pm$ 4~keV & 29.4 k & $14.1 \pm 0.2$ \\
\noalign{\smallskip}\hline
\end{tabular}
\end{table}

\begin{figure}
	\includegraphics{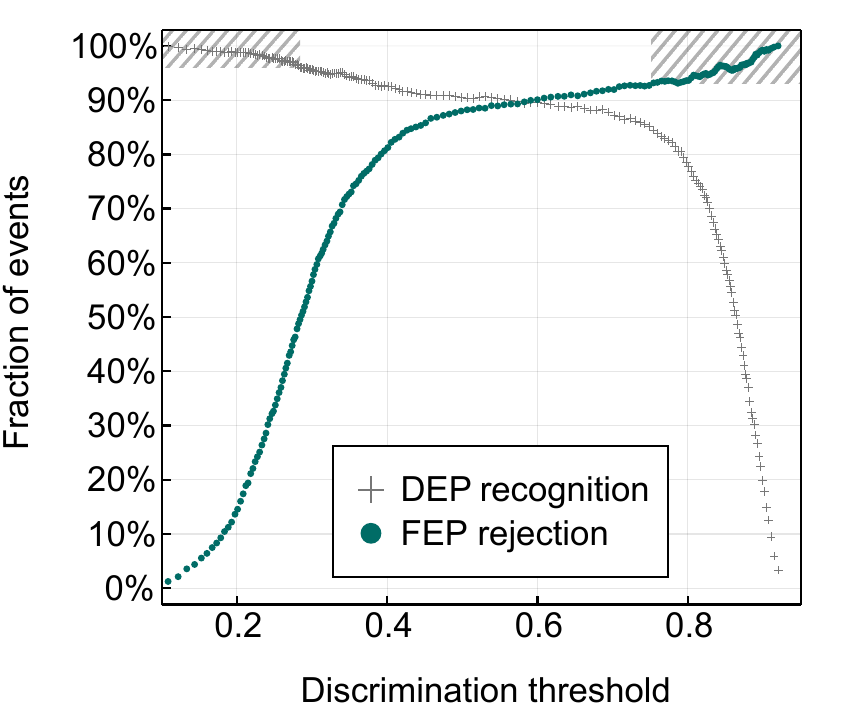}
	\caption{SS event recognition and MS event rejection efficiencies as a function of the discrimination threshold on the classifier output for the DEP (mostly SS) and the FEP (mostly MS). Inside the highlighted areas, more events are accepted or rejected than the peak contains of the associated class.}
	\label{fig:efficiencies}
\end{figure}

Figure~\ref{fig:efficiencies} shows how the survival fraction of DEP events and the rejection of FEP events vary depending on the discrimination threshold.
All survival fractions are obtained from a binned fit of a linear background plus a Gaussian to the peaks in the energy spectrum.
From our MC simulations, we know that 93\% of true DEP events (excluding Compton background) are SS and 96\% of true FEP events are MS.
Higher recognition or rejection rates necessarily accept or reject too many events. These areas are highlighted in Fig~\ref{fig:efficiencies}.

For our classifier, a discrimination threshold of 0.6 yields maximum class separation efficiency with 90\% FEP rejection at 90\% DEP recognition.
At this threshold, 90k of the 232k events in the test dataset which have energy depositions in segment~1 are classified as SS and 141k events as MS. 
In the same dataset, 100k events are labeled as single-segment and 132k as multi-segment.
Table~\ref{tab:correlation} shows the number of events classified as SS and MS for both single-segment and multi-segment events.

\begin{table}
\centering
\caption{Classifications by the neural network (discrimination threshold of 0.6) of single-segment and multi-segment events.}
\label{tab:correlation}
\begin{tabular}{lll}
\hline\noalign{\smallskip}
 & Single-site & Multi-site \\
\noalign{\smallskip}\hline\noalign{\smallskip}
Single-segment & 65 k & 35 k \\
Multi-segment & 26 k & 106 k \\
\noalign{\smallskip}\hline
\end{tabular}
\end{table}

Of all single-segment events, 65\% are classified as SS and 80\% of the multi-segment events are classified as MS.
The relatively high fraction of single-segment events that are classified as multi-site is not surprising because gammas can cause multiple energy depositions within one segment, resulting in single-segment MS events.

\section{Discussion}
\label{sec:discussion}

\begin{figure}
	\includegraphics{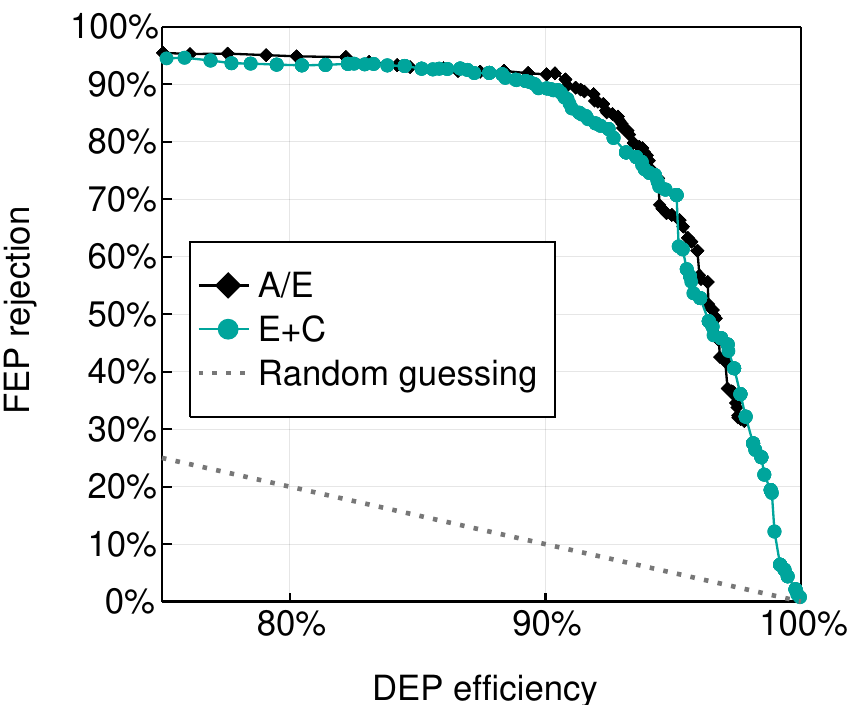}
	\caption{Rejection power of the combined encoder+classifier network (E+C) compared to the $A/E$ discrimination algorithm.}
	\label{fig:roc}
\end{figure}

In order to assess the overall discrimination performance of our method,
we compare it to the performance of the $A/E$ technique currently employed for BEGe type detectors by $0\nu\beta\beta$ experiments (see section~\ref{sec:introduction}).
The $A/E$ survival fractions are calculated with the same fitting method, described in section~\ref{sec:verification}.
Figure~\ref{fig:roc} compares the rejection power of the combined E+C network with the power of the $A/E$ method for equal DEP survival fractions.

Despite the similarity of the two performance curves, the classifiers are fundamentally different and often do not agree in their classifications.
While $A/E$ is based on a single parameter (the maximum current divided by the total deposited energy), the E+C network takes the whole waveform into consideration.
Using a threshold chosen to result in a 90\% DEP survival fraction, the two methods assign different classes to about 10\% of all events.
The two methods can therefore be regarded as complementary.
The events classified as SS by the E+C network and as MS by $A/E$ account for about three quarters of these events and 52\% of them are labeled as single-segment.
On the other hand, 79\% of the events classified as SS by $A/E$ but classified as MS by E+C are multi-segment, and therefore almost certain to be true MS events.
This demonstrates that the E+C network can identify certain types of MS events that are incorrectly classified as SS by the $A/E$ method.

\begin{figure}
	\includegraphics{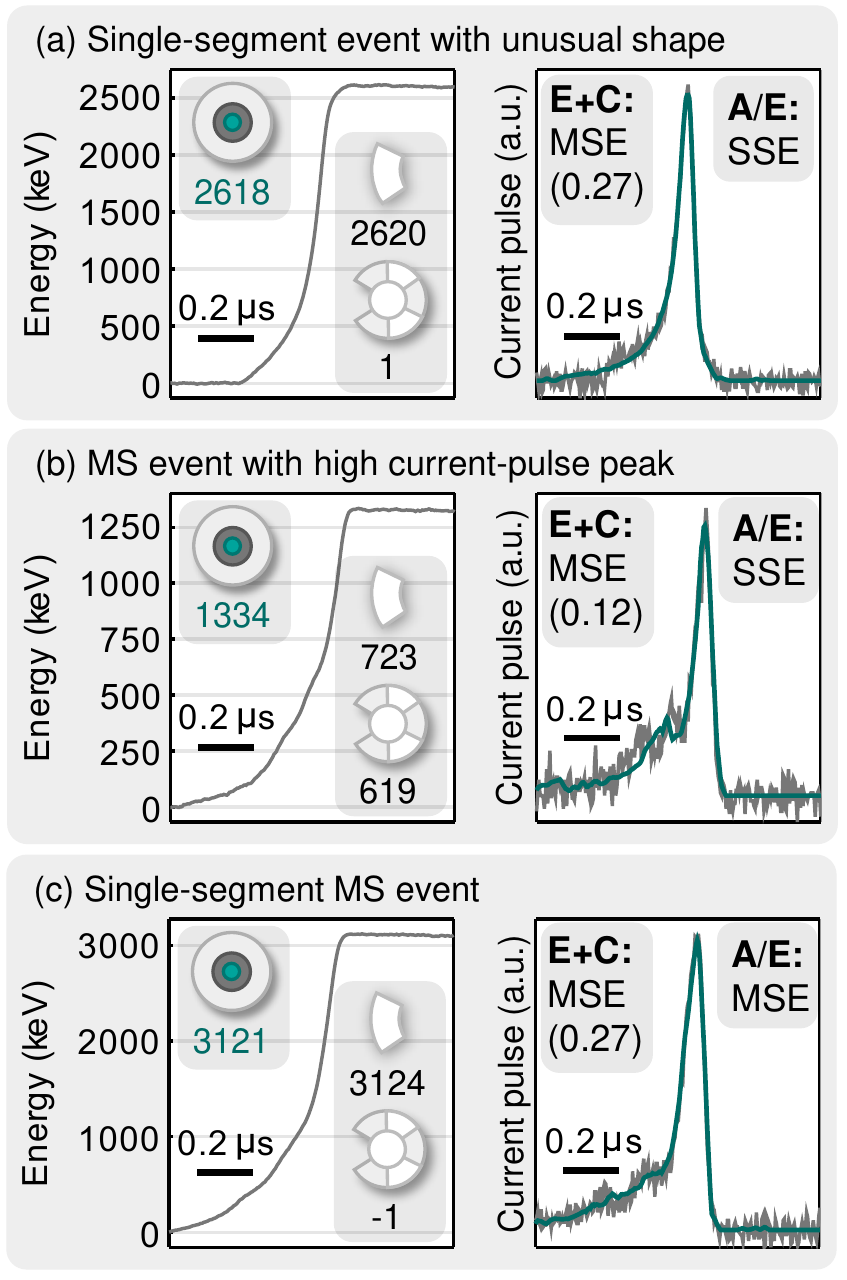}
	\caption{Example pulses illustrating different classifications of the $A/E$ and the encoder+classifier (E+C) methods. Left: The charge-signal waveforms, also indicating the core energy (in keV), the energy in segment~1 and the energy in other segments. Right: The preprocessed current-signal waveforms (gray curves) with their autoencoder reconstructions (green curves), indicating the event classifications and output value of the E+C network and the class predicted by the $A/E$ method.}
	\label{fig:disagreement}
\end{figure}

These differences in classification become clear when examining specific waveforms.
Figure~\ref{fig:disagreement}a shows a 2618~keV event that deposited all energy in segment~1. The rise in the charge-pulse starts more abruptly than with most SS events and the current-waveform shows larger-than-normal noise fluctuation just before its peak. While $A/E$ classifies this as SS, the E+C network classifies it as MS.

The waveform of the multi-segment 1334~keV event in Fig.~\ref{fig:disagreement}b, classified in the same manner, looks similar except for multiple small substructures in the signal. However, as only about half of the energy is deposited in segment~1, it is almost certainly multi-site.
Part of the energy may be deposited close to the contact, causing a high peak in the current waveform which in turn causes $A/E$ to misclassify it as SS.
Our test dataset contains about 4500 events of this type.

Figure~\ref{fig:disagreement}c shows an event with all energy deposited in segment~1. Events at such high energies are almost exclusively coincidence events, i.e. events caused by two or more gamma rays depositing energy at the same time.
It can also clearly be identified as MS from the multiple-peak structure in the current-pulse and both $A/E$ and E+C classify it as such.
This demonstrates that segment information alone is not sufficient to detect all MS events.

All machine learning techniques heavily rely on the quality of their training data since no prior knowledge of physical processes is assumed.
It is therefore remarkable that the E+C network matches the performance of the physics-based $A/E$ method despite our small and impure training dataset.

The electronics of our detector system has a relatively low noise level compared to larger scale experiments.
It has been demonstrated, using \gerda data, that our classification is robust in the presence of different types of noise or variation in waveform shape due to changes of detector or amplifier characteristics over time~\cite{MasterThesis}.
This robustness stems from the fact that the classifier only depends on the extracted features equivalent to the denoised waveform.
This is an advantage over the $A/E$ method, of which the classification performance directly depends on the noise level: The low noise level of the data used in this work can be seen as a best-case scenario for $A/E$.

\section{Conclusions}
\label{sec:conclusions}

We have demonstrated the use of two different deep-learning based neural networks in combination to achieve state-of-the-art discrimination performance for single-site / multi-site recognition for germanium detectors.
By splitting the discrimination method into two independent stages, a feature extraction and a classification stage (E+C), only a small subset of all training data needs to be labeled.
The first stage is a feature-extraction performed by an autoencoder. It drastically reduces the dimensionality of the data while retaining the essential characteristics of the waveform.
This network is trained in an unsupervised fashion, so no class labels are required.
Using only seven feature parameters, the waveforms are approximated with sufficient accuracy.

The classification network that operates on the extracted feature vectors has been shown to be competitive in discrimination performance with the widely used $A/E$ algorithm, despite being trained on a small and impure dataset.

Our method is currently limited by the volume and purity of the training data.
We are working on an accurate pulse shape simulation that could provide arbitrary amounts of high-purity synthetic training data.
Assuming the simulation reaches a sufficient level of accuracy, one might also be consider to train the autoencoder on measured data and the classifier on simulated data.
This would combine the properties of a real detector system, especially its noise characteristics, with a classifier trained on synthetic data with pure labels.

Due to their fundamentally different approach it may also be profitable to combine our E+C method with the $A/E$ method, this may result in an even more powerful background rejection scheme.
One way to achieve this would be to adjust the classification thresholds and only accept an event if both methods classify it as SS.
As has been shown, the $A/E$ technique fails to reject certain types of multi-site events, e.g. events with energy depositions close to the detector contact.
Since the E+C network has access to the whole waveform shape, it can identify them like any other MS event.

The E+C method presented here is powerful enough to encode and classify events in an automated fashion for a large number and different types of detectors without manual corrections~\cite{MasterThesis}.
It has the potential to be a valuable background rejection technique for the next generation of $0\nu\beta\beta$ decay experiments, e.g. \legend~\cite{Legend}.

\begin{acknowledgements}
We would like to thank Iris Abt for her valuable comments to this manuscript.
\end{acknowledgements}

\par\addvspace{17pt}
\small\rmfamily{The data used in this paper can be requested from the authors.}

\bibliographystyle{spphys}       
\bibliography{bibliography}

\end{document}